\documentstyle[epsfig]{article}
\begin{document}
\begin{center}
  {\large\bf Extremal Entanglement For Triqubit Pure States}
\vskip .6cm {\normalsize Feng Pan,$^{1,2}$
Dan Liu$^{1}$, Guoying Lu$^{1}$, and J. P. Draayer$^{2}$ }
\vskip .2cm {\small
$^{1}$Department of Physics, Liaoning Normal University, Dalian
116029, P. R. China\vskip .1cm
$^{2}$Department of Physics and Astronomy,
Louisiana State University, Baton Rouge, LA 70803-4001}
\end{center}
\vskip .5cm
  {\bf Abstract:~}\normalsize A complete analysis of entangled triqubit
pure states is carried out based on a new simple entanglement measure.
An analysis of all possible extremally entangled pure triqubit states
with up to eight terms is shown to reduce, with the help of SLOCC
transformations, to three distinct types. The analysis presented
are most helpful for finding different entanglement types
in multipartite pure state systems.
\vskip .5cm
Entanglement is a fundamental concept that underpins
quantum information and computation.$^{[1-3]}$
As a consequence, the quantification of entanglement emerges as
a central challenge.
Many authors have contributed to this topic,$^{[4-23]}$
among which the basic requirements for entanglement
measures proposed in [4] provide with guidelines for
its definition. In [6], Bennett {\it et al}
defined stochastic local operations
and classical communication (SLOCC) based
on the concept of local operations
assisted with classical communication (LOCC).
D\"{u}r {\it et al} applied such an operation
to a triqubit pure state system and found that triqubit
states can be entangled {\it at least } in two
inequivalent ways,$^{[7]}$ namely, in the GHZ form$^{[24]}$ or the
W form.$^{[7]}$ In this Letter, we will use a recently proposed
entanglement measure$^{[23]}$ for $N$-qubit pure states to analyze all
extremally entangled triqubit pure states with the
constrained maximization, and to see
whether there are other inequivalent
types of entanglement.

In the following, based on the method used in [7],
only entanglement properties of a single copy of a state
will be considered. Therefore, asymptotic properties
will not be discussed. At single copy level,
it is well known that
two pure states can always be transformed with certainty
from each other by means of LOCC if and only if they are
related by Local Unitary transformation (LU).$^{[6]}$
However, even in the simplest bipartite cases,
entangled states  are not  always related by
LU, and continuous parameters are needed to label all
equivalence classes. Hence, it seems that one needs to deal with
infinitely many kinds of entanglement. Fortunately, such
arbitrariness has been overcome with the help of SLOCC.$^{[7]}$

\vskip .3cm
According to [23], for a genuine entangled $N$-qubit pure state
$\Psi$, the measure can be defined by
$$E(\Psi)=
\left\{
\begin{array}{cc}
{1\over{N}}\sum^{N}_{i=1}
{S_{i}}&{{\rm if} ~S_i\neq 0~\forall~i},\\
{}\\
0~~~~~~~~~&{\rm otherwise,}\\
\end{array}
\right.\eqno(1)$$
where $S_i=-{\rm Tr}[\left(\rho_{\Psi}\right)_i
\log_2{\left(\rho_{\Psi}\right)_i}]$ is the reduced von Neumann entropy
for the $i$-th particle only with the other $N-1$ particles traced out,
and $\left(\rho_{\Psi}\right)_i$ is the corresponding reduced density
matrix. It can be verified that the state $\Psi$ is partially
separable when one of the reduced von Neumann entropies $S_i$ is zero. In
such cases the state $\Psi$ is not a genuine entangled $N$-qubit state.
Furthermore, definition (1) is invariant under LU,
which is equivalent to LOCC for pure state
system. Eq. (1) will be our unique benchmark
for the degree of entanglement of $N$-qubit
pure states. Using (1), we have successfully verified that there is only
one type of extremal (maximal) entanglement for biqubit system, which is
equivalent to the Bell type.$^{[25]}$
\vskip .3cm
The basis vectors of a triqubit system are
denoted by
$$\begin{array}{lllll}
& &
\{W_1=\vert000\rangle, W_2=\vert110\rangle,
  W_3=\vert101\rangle,W_4=\vert011\rangle,\\
& & \bar{W_1}=\vert111\rangle,\bar{W_2}=\vert001\rangle,
\bar{W_3}=\vert010\rangle,\bar{W_4}=\vert100\rangle\}.
  \end{array}\eqno(2)$$
\noindent In order to study all possible forms of entangled
triqubit states, entangled states are classified according to the
number of terms involved in their expressions
in terms of a linear combination of basis vectors given in (2).
This means there will be up to eight terms in their expressions.
Then, maximization constrained by the normalization condition
for the entanglement measure (1) is performed
to find the corresponding parameters
and phase factors. In this way, all extremally entangled triqubit pure
states can be obtained. It will be shown that values of the measure for
different extremally entangled states are all different, and
these extremally entangled forms can thus be recognized
as the different types of entanglement for triqubit system
since they can not be transformed into other forms by  SLOCC.

\vskip .3cm
When a state $\Psi$ is a linear combination
of two terms $W_i$ and $W_j$ or $\bar{W}_{i}$ and $\bar{W_j}$,
where $i,~j\in P$ ($i\neq j$), and $P=\{1,2,3,4\}$,
$\Psi$ is a partially separable state.
There are $12$ such combinations.
It can easily be verified that $E(\Psi)=0$ in such cases
according to the measure defined in (1).
In such cases, one of reduced von Neumann entropy
$S_{i}$ ($i=1,2,3$) is zero.
As an example, When $\Psi=aW_{1}+be^{i\alpha}W_{2}$
with constraint $|a|^2+|b|^2=1$, one obtains

$$\left(\rho_{\Psi}\right)_{1}=\left(\rho_{\Psi}\right)_{2}
=\left(\matrix{|a|^2 &\cr  &|b|^2}\right),~~
\left(\rho_{\Psi}\right)_{3}=\left(\matrix{1 &\cr
&0}\right).\eqno(3)$$
Therefore, in this case, generally $S_{1}=S_{2}\neq 0$, while
$S_{3}=0$. According to (1), the measure  $E(\Psi)=0$ since
one of the reduced von Neumann entropy, $S_{3}$ is zero.
Analysis on the other $11$ combinations are similar,
which lead to the same conclusion.
Similarly, when $\Psi$ is a linear combination
of two terms $W_i$ and $\bar{W_j}$ with $i\neq j$,
$\Psi$ is a fully separable state, of which
$E(\Psi)=0$ is obvious.
There are also $12$ of such linear combinations.
$\Psi$  will be an entangled state
only when it is a linear combination
of $W_i$ and $\bar {W_i}$.
In such cases, one may write $\Psi=
a~W_j+b e^{i\alpha}~\bar{W_j}$,
where $a,~b$ are real, and $\alpha$ is a relative
phase. Then, to maximize its measure (1) with
the constraint $a^2+b^2=1$, one can find
parameters for the corresponding extremal cases.
It can be
verified easily that $E=E_{\rm max}=1$ when
$\vert a\vert=\vert b\vert=1/\sqrt{2}$ in such cases,
and there is no restriction on relative phase.
Such exremally (maximally) entangled two-term states are nothing
but the GHZ states.
\vskip .2cm
When a state $\Psi$ is a linear combination
of three terms shown in (1),
there are $(^8_3)=56$ forms of $\Psi$ in total, among which
$24$ of them are partially separable, and the remaining
$32$ of them are genuine entangled states.
When $\Psi=a~ W_i+be^{i\alpha}~\bar{W_i}+ce^{i\beta}~ W_j$ or $\Psi
=a~W_i+be^{i\alpha}~\bar{W_i}+ce^{i\beta}~\bar{ W_j}$,
where $i,~j\in P$ ($i\neq j$),
$a$,~$b$, $c$, $\alpha$ and $\beta$ are real,
there are $24$ such combinations.
In such cases, one can verify that no extremal value of
$E(\Psi)$ exists if coefficients $a$,~$b$,~$c$ are all non-zero.
When a state  $\Psi$ is a linear combination of ($ W_i, W_j,W_q$)
or that of ($\bar{W_i},\bar{W_j},\bar{W_q}$),
where $i$,~$j$,~$q\in P$ with $i\neq j\neq q$, the state is a genuine
entangled state. Similar to the GHZ states, one can prove that extremal
value of $E(\Psi)$ is $0.918296$ for such states when
the absolute value of all the expansion coefficients equal to
$1/\sqrt{3}$. The result is also independent of the relative
phases. These entangled states are nothing but the W
states.
\vskip .2cm

When a state $\Psi$ is a linear combination
of four terms shown in (1), there are $70$ forms in total, among which
$6$ of them are partially separable, while the remaining
$64$ of them are  genuine entangled states. They are classified
into $5$ types shown in Table 1, in which Yes (No) listed
in the last column refers to the fact that there exists (does not exist)
an extremal value of $E(\Psi)$ for the corresponding state with all
expansion coefficients nonzero.

\newpage

\begin{center}
\begin{center}{{\bf Table 1.}~~~Classification of the entangled states
with four terms}
\end{center}\vskip .4cm
{
\begin{small}
\begin{tabular}{cccccc}\hline\hline
{Type}&{Terms involved}&{Total Number of Forms}&{Extremal entanglement}\\
  \hline
  {}\\
  {Type I$_{4}$}&{$ W_i, W_j,
\bar{W_i}, \bar{W_j}$}
&{$6$}&{Yes}\\
&{($i,j\in P,~i\neq j$)}&&\\

  {Type II$_{4}$}&{$ W_i, W_j,
  {W_t},{W_q}$ or}
&{$2$}&{Yes}\\
&{$\bar{W_i},\bar{ W_j},
\bar {W_t},\bar{W_q}$}&&\\
&{($i,~j,~t,~q\in P,~i\neq j\neq t\neq q$)}&&\\

  {Type III$_{4}$}&{$W_i, W_j, {W_t},\bar{W_i}$ or}
&{$24$}&{Yes}\\
&{$ \bar{W_i},\bar{ W_j},
\bar {W_t},{W_i}$}&&\\
&{($i,j,t\in P,~i\neq j\neq t$)}&&\\

  {Type IV$_{4}$}&{$ W_i, W_j,\bar{W_i},\bar{W_q}$}
&{$24$}&{No}\\
&{($i,j,q\in P,~i\neq j\neq q$)}&&\\

  {Type V$_{4}$}&{$ W_i,W_j,{W_t},\bar{W_q}$ or}
&{$8$}&{No}\\
&{$\bar{W_i},\bar{ W_j},\bar {W_t},{W_q}$}&&\\
&{($i,j,t,q\in P,i\neq j\neq t\not= q$)}&&\\
{}\\
\hline\hline
\end{tabular}\end{small}
  }
  \end{center}

For the Type I$_{4}$ case,
$\Psi=a~\bar{W_{i}}+be^{i\alpha} {W_{i}}+
ce^{i\beta}W_{j}+de^{i\gamma}\bar{W_{j}}$,
where $i\neq j$, and $a,~b,~c,~d$, $\alpha,~\beta,~\gamma$ are real with
constraint $a^2+b^2+c^2+d^2=1$.
One can verify that $E(\Psi)=1$ when
$\vert a\vert=\vert b\vert,
~\vert c\vert=\vert d\vert,~\theta=\alpha-\beta-\gamma=(2n+1)\pi$
with integer $n$.
Such states are equivalent to the GHZ type states after LU transformations.
For Type II$_{4}$ case, $\Psi=a W_{i}+be^{i\alpha} W_{j}+
ce^{i\beta} W_{t}+de^{i\gamma} W_{q}$,
where $i\neq j\neq t\neq q$.
When $\vert a\vert=\vert b\vert=\vert c\vert=\vert d\vert={1\over2}$
for arbitrary $\alpha,~\beta,~\gamma$,
the corresponding states are also equivalent to the GHZ type states. For
Type III$_{4}$ case, $\Psi=a W_{j}+ be^{i\alpha} W_{t}+ce^{i\beta} W_i
+de^{i\gamma}\bar{W}_{i}$, where $i\neq j\neq t$.
In this case, the reduced density matrices are
$$\rho_A=\left(\begin{array}{cc}c^2&0\\
0&a^2+b^2+d^2\end{array}\right),
~\rho_B=\left(\begin{array}{cc}b^2+d^2&ade^{i\gamma}\\
ade^{-i\gamma}&a^2+c^2\end{array}\right),$$
$$\rho_C=\left(\begin{array}{cc}a^2+d^2&bde^{i\gamma-i\alpha}\\
bde^{-i\gamma+i\alpha}&b^2+c^2\end{array}\right).\eqno(4)$$
The corresponding entanglement measure is
$$\begin{array}{c}
E(\Psi)={1\over3}\{-c^2\log_2{c^2}-(a^2+b^2+d^2)\log_2{(a^2+b^2+d^2)}-
{1\over2}(1-u)\log_2{[{1\over2}(1-u)]}\\
-{1\over2}(1+u)\log_2{[{1\over2}(1+u)]}
-{1\over2}(1-v)\log_2{[{1\over2}(1-v)]}-
{1\over2}(1+v)\log_2{[{1\over2}(1+v)]}\},
\end{array}\eqno(5)$$
where
$$u=\sqrt{1-4(a^2b^2+a^2c^2+c^2d^2)},~~
v=\sqrt{1-4(a^2b^2+b^2c^2+c^2d^2)}.\eqno(6)$$

We found that there are three extremal values for the measure
with $E_{1}(\Psi)=1$ when $\vert a\vert=\vert b\vert=0$,
$\vert c\vert=\vert d\vert=1/\sqrt{2}$,
and $E_{2}(\Psi)=0.918296$ when $\vert a\vert=\vert b\vert=\vert c\vert=1/\sqrt{3}$,
$\vert d\vert=0$, and $E_{3}(\Psi)=0.893295$ when
$\vert a\vert=\vert b\vert=0.462175,~\vert c\vert=0.653614,~
\vert d\vert=0.381546$, respectively.
There is also no restriction on relative phases.
The first and the second cases
clearly correspond to the GHZ and W
types, respectively. While the third case with
$E_{3}(\Psi)=0.893295$ has not been reported
previously. In order to study whether this type
of entangled states is equivalent to GHZ or W types,
in the following, we will briefly review
the SLOCC used in [7]~:
States $\Psi$ and $\Phi$ are equivalent under SLOCC if
an invertible local operator (ILO)
relating them exists, which is denoted as
$Q_{A}\otimes Q_B\otimes Q_C$. Typically, these ILOs
are elements of the complex general linear group
$GL_{A}(2,c)\otimes GL_{B}(2,c)\otimes
GL_{C}(2,c)$  with each copy operates on the
corresponding local basis.
\vskip .2cm
In the following, we prove that the Type III$_{4}$ entangled states
is inequivalent to the GHZ and W types under the SLOCC. To do so, we take
$$\Psi=x_{1}W_{2}+x_{2}W_{3}+x_{3}W_{4}+x_{4}\bar{W}_{4}\eqno(7)$$
as a representative of Type III$_{4}$ entangled states
since the proof for other cases is similar,
where $x_{i}$ ($i=1,\cdots,4$) are arbitrary nonzero complex numbers.
Let $Q_{A}$, $Q_{B}$, and $Q_{C}$ be ILOs
for particles $A$, $B$, and $C$, respectively.
First, we prove that (7) is inequivalent to a GHZ type state.
Notice that Eq. (7) can be rewritten as
$$\Psi=\vert 1\rangle_{A}\left(
x_{1}\vert 10\rangle_{BC}+x_{2}\vert 01\rangle_{BC}\right)+\left(
x_{3}\vert 011\rangle_{ABC}+x_{4}\vert 100\rangle_{ABC}\right).\eqno(8)$$
The first term in (8) is a product of a Bell type state and a single
particle state, while  the second term is a GHZ type state.
After  $Q_{A}\otimes Q_{B}\otimes Q_{C}$ transformation,
the first term will remain a product of a single
particle state and a Bell type state, while the second term
will always remain a GHZ type state. Therefore, (7) is
inequivalent to GHZ type state if $x_{1}$ and $x_{2}$ are all
non-zero. It is clear that there is a phase transition from
the Type III$_{4}$ to the GHZ type when the parameters $x_{1}$
and $x_{2}$ tend to zero. Such a transition belongs to the first
class phase transition. Similarly, the Type III$_{4}$ entangled
state becomes separable when $x_{3}$ and $x_{4}$ tend to zero.
Second, we prove that (7) is also inequivalent to W type state.
Eq. (7) can also be written as
$$\Psi=\left(
x_{1}\vert 110\rangle_{ABC}+x_{2}\vert 101\rangle_{ABC}+x_{3}\vert
011\rangle_{ABC}\right)+
x_{4}\vert 100\rangle_{ABC}.\eqno(9)$$
The first term in (9) belongs to a W type state, while the second term
is a product of three single-particle states.
After arbitrary $Q_{A}\otimes Q_{B}\otimes Q_{C}$ transformation,
the first term will remain a W type state, while the second
term will always a product of three single-particle states.
Hence, the Type III$_{4}$ entangled
state is also inequivalent to a W Type if $x_{4}$ is nonzero.
Similarly, the Type III$_{4}$  entangled
state will become a W type state only when $x_{4}$ tends to
zero. Furthermore, it can be shown by a simple analysis
that invertible transformation from (7)
with all parameters nonzero to either GHZ
or W state does not exist.
Therefore, for nonzero $x_{i}$ ($i=1,\cdots,4$),
the Type III$_{4}$ is a new type of entanglement in triqubit
pure system, which is inequivalent to the GHZ and W types.
This result is not so surprising since
the Type III$_{4}$ states were not studied in [7].
There is no extremal value of the measure
for types IV$_{4}$ and V$_{4}$ with all expansion coefficients
nonzero.

\vskip .2cm

When a state $\Psi$ is a linear combination
of five terms, there are $(^8_5)=56$ of them, which can be
classified into three types.
If $\Psi$ is a linear combination of
($ W_1, W_2,
{W_3},{W_4},\bar{W_i}$)
or ($\bar{W_1},\bar{W_2},
\bar{W_3},$ $\bar{W_4},{W_i}$), where $i\in P$,
it is called Type I$_{5}$.
If $\Psi$ is a linear combination of
($W_i, W_j, {W_q},\bar{W_i},\bar{W_j}$)
or ($\bar{W_i},\bar{W_j},\bar{W_q},{W_i},{W_j}$),
where $i,~j,~q\in P$, it is called Type II$_{5}$.
If $\Psi$ is a linear combination of ($ W_i, W_j,
{W_q},\bar{W_i},\bar{W_t}$)
or ($\bar{W_i},\bar{W_j},
\bar{W_q},{W_i},$ ${W_t}$), where
$i,~j,~q,~t\in P,~i\neq j\neq t\neq q$, it is called Type III$_{5}$.
For the Type I$_{5}$ cases, $\Psi=a W_1+be^{i\alpha} W_2
+ce^{i\beta} W_3+de^{i\gamma} W_4
+fe^{i\sigma} \bar{W_i}$, where $a,~b,~c,~d,~f$,
$\alpha,~\beta,~\gamma,~\sigma$ are real.
When $\vert a\vert={{2\over3}},~\vert
b\vert=\vert c\vert=\vert d\vert={{1\over3}}$, and $\vert
f\vert={{\sqrt{2}\over3}}$ with arbitrary phases,
there exists a extremal value of the measure
with $E(\Psi)=0.918296$. It can be verified that such states are
equivalent to W type states with respect to SLOCC.
For the Type II$_{5}$ and Type III$_{5}$ cases, we found that there is
no extremal value of the measure exists with all expansion
coefficients nonzero.
Similar analysis for states with six, seven, and eight terms
were also carried out. After tedious computations,
we did not find any other new type of extremally entangled states.
The results are consistent with the conclusion made in [10] and [22]
that entangled triqubit pure states can always be expressed as
a linear combination of five appropriate terms chosen from (2).

\vskip .2cm

In summary, by using the entanglement measure (1),
a complete analysis for entangled triqubit pure states
has been carried out.
Three types of extremally entangled triqubit pure states
have been identified by using the constrained maximization.
The extremal values of
these three types of entanglement are  $1$, $0.918296$,
$0.893295$, respectively, which show that
the GHZ type state with

$$\vert {\rm GHZ}\rangle={1\over{\sqrt{2}}}(\vert 000\rangle+
\vert 111\rangle)\eqno(10a)$$
and the W type with

$$\vert {\rm W}\rangle={1\over{\sqrt{3}}}(\vert 100\rangle+
\vert 010\rangle+\vert 001\rangle)\eqno(10b)$$
are all the special cases of the Type III$_{4}$ states.
Similarly, the type III$_{4}$ states can always be transformed
by SLOCC to the following form

$$\vert {\rm III}_{4}\rangle=a\vert 110\rangle+b\vert 101\rangle+c\vert 011\rangle+
d\vert 100\rangle\eqno(10c)$$
with $\vert a\vert=\vert b\vert=0.462175,~\vert c\vert=0.653614,~
\vert d\vert=0.381546$, respectively.
These three types of entangled states cannot be
transformed from one type into another type by SLOCC,
which, therefore, are recognized to be
all inequivalent types of entanglement.
Graphical descriptions of these three
types of tripartite entanglement are shown
in Fig. 1. The relations among these three types of
entanglement are shown in Fig. 2.

\vskip .3cm

We also found that the entanglement
measure defined by (1) is effective in classifying
different genuinely entangled tripartite pure states.
The most important conclusion is that the number of
basic ways of entanglement equals to  the number of extremally
entangled types. Therefore, extremal
entanglement is a necessary condition
in finding different ways of entanglement
in multipartite pure state systems under the SLOCC.
The analysis can be extended to  other
multipartite pure state systems in a straightforward manner.

\vskip .2cm
Support from the U.S. National Science Foundation
(0140300), the Southeastern Universitites Research
Association, the Natural Science Foundation of China
(10175031), the Natural Science Foundation of Liaoning
Province (2001101053), the Education Department
of Liaoning Province (202122024), and the LSU--LNNU joint
research program is acknowledged.

\vskip .5cm

\def\HT{\bf\relax}
\def\REF#1{\small\par\hangindent\parindent\indent\llap{#1\enspace}\ignorespaces}

\section*{Reference}

\noindent\REF{[1]}  C. H. Bennett and D. P. DiVincenzo, Nature (London)
{\bf 404}, 247(2000)

\REF{[2]}  M. A. Nielsen and I. L. Chuang,
Quantum Computation and Quantum Information
(Cambridge University Press, Cambridge, England, 2000)

\REF{[3]}  A. Rauschenbeutel, G. Nogues, and S. Osnaghi, Science {\bf
288}, 2024(2000)

\REF{[4]} M. Horodecki, P. Horodecki, and R. Horodecki,
Phys. Rev. Lett. {\bf 84}, 2014 (2000)

\REF{[5]} V. Vedral, M. B. Plenio, M. A. Rippin, and P. L. Knight,
Phys. Rev. Lett. {\bf 78}, 2275 (1997)

\REF{[6]} C. H. Bennett, S. Popescu, D. Rohrlich, J. A. Smolin, and
A. V. Thapliyal,
Phys. Rev. A {\bf 63}, 012307(2001)

\REF{[7]} W. D\"{u}r, G. Vidal, and J. I. Cirac,
Phys. Rev. A {\bf 62}, 062314 (2000)

\REF{[8]} A. V. Thapliyal, Phys.  Rev. A {\bf 59}, 3336 (1999)

\REF{[9]} S. J. Wu and Y. D. Zhang, Phys. Rev.  A {\bf 63}, 012308 (2001)

\REF{[10]} A. Acin, A. Andrianov, L. Costa, E. Jan\'{e},
  J. I. Latorre, and R. Tarrach, Phys. Rev. Lett. {\bf 85}, 1560 (2000)

\REF{[11]} N. Linden and S. Popescu,  Fortsch. Phys. {\bf 46}, 567 (1998)

\REF{[12]} A. Sudbery, J. Phys. A {\bf 34}, 643 (2001)

\REF{[13]} V. Coffman, J. Kundu, and W. K. Wootters,
  Phys. Rev. A {\bf 61}, 052306 (2000)

\REF{[14]} C. H. Bennett, H. J. Bernstein, S. Popescu, and B. Schumacher,
Phys. Rev. A {\bf 53}, 2046 (1996)

\REF{[15]} L. L. S\'{a}nchez-Soto, J. Delgado, A. B. Klimov,
  and G.Bj\"{o}rk, Phys. Rev. A {\bf 66}, 042112 (2002)

\REF{[16]} A. K. Rajagopal and R. W. Rendell, Phys. Rev. A65, 032328(2002)

\REF{[17]} A. K. Rajagopal and R. W. Rendell, Phys. Rev. A66, 022104(2002)

\REF{[18]} H. Heydari, G. Bj\"{o}rk, L. L. S\'{a}nchez-Soto,
  Phys. Rev. A {\bf 68}, 062314 (2003)

\REF{[19]} H. Heydari, and G. Bj\"{o}rk, arXiv:quant-ph/0401129

\REF{[20]} K. Audenaert, M. B. Plenio, and J. Eisert, Phys. Rev. Lett.
{\bf 90}, 027901 (2003)

\REF{[21]} Sixia Yu, Zeng-Bing Chen, Jian-Wei Pan, and Yong-De Zhang,
Phys. Rev. Lett. {\bf 90}, 080401 (2003)

\REF{[22]} M. H. Partovi, Phys. Rev. Lett. {\bf 92}, 077904 (2004)

\REF{[23]} Feng Pan, Dan Liu, Guoying Lu, and J. P. Draayer, Int. J.
Theor. Phys. {\bf 43}, 1241 (2004)

\REF{[24]} D. M. Greenberger, M. A. Horne, and A. Zeilinger, in Bell's
Theorem, Quantum Theory, and Conceptions of the Universe, edited by M.
Kafatos (Kluwer, Dordrecht, 1989) 69

\REF{[25]} J. S. Bell, Physics (Long Island City, N. Y.) {\bf 1}, 195
(1964)

\newpage

\begin{figure}[tbp]
\centerline{\hbox{\epsfig{file=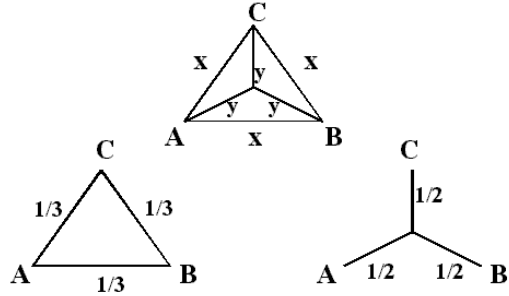, width=11cm}}}
\caption{Graphical descriptions of three
types of entanglement. The one in the upper middle represents
the type III$_{4}$ with ${\rm x}=|ab|/3={|bc|\over{{3\sqrt{2}}}}={|ac|\over{{3\sqrt{2}}}}$,
and ${\rm y}={|cd|\over{{2\sqrt{3}}}}$, where
$|a|$, $|b|$, $|c$, and $|d|$ are given after Eq. (10c).
The lower left one represents
the W type, in which the value besides each edge is $x_{1}x_{2}=
x_{2}x_{3}=x_{1}x_{3}={1\over{3}}$ with
$x_{1}=x_{2}=x_{3}=1/\sqrt{3}$, and the Y
type connection within the triangle
disappears since $x_{4}=0$ in this case.
The lower right one represents the GHZ type, in which the value
besides each line is $x_{3}x_{4}=1/2$ with
$x_{3}=x_{4}=1/\sqrt{2}$, and the triangle disappears
since $x_{1}=x_{2}=0$ in this case.}
\end{figure}

\begin{figure}[tbp]
\centerline{\hbox{\epsfig{file=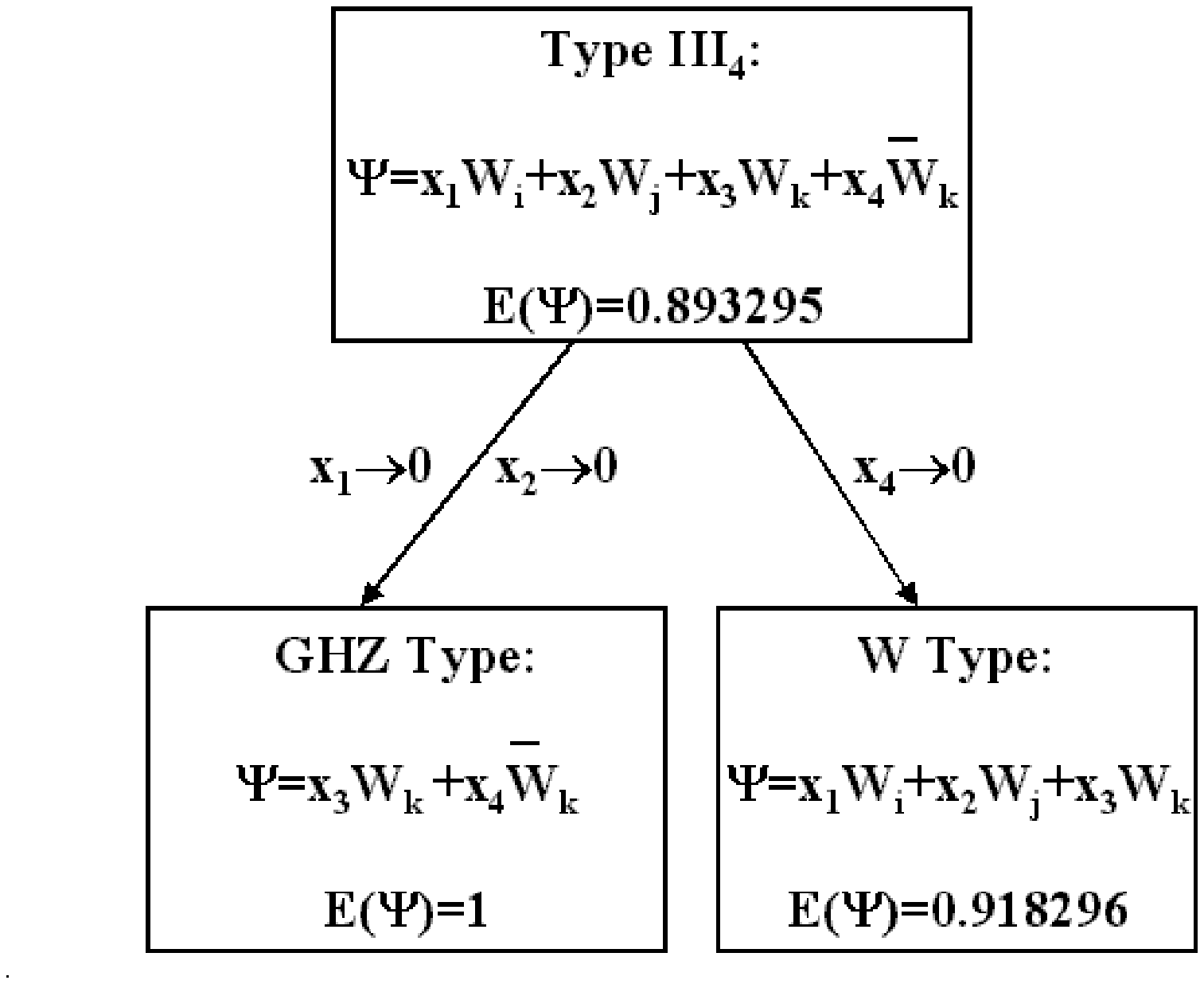, width=9cm}}}
\caption{Relations among three inequivalent
types of entanglement, where $i,~j,~k\in P$ with
$i\neq j\neq k$, and the measures listed are the
corresponding extremal values according to (2).
These three types of entangled states can always be
transformed under SLOCC into the corresponding forms shown by Eq. (10).
}
\end{figure}

\end{document}